\documentclass{PoS}
\usepackage{subfigure}
\usepackage{xspace}
\newcommand{\mt     }{\ensuremath{m_{\mathrm{top}}}\xspace}
\newcommand{\mtr    }{\ensuremath{m_{\mathrm{top}}^{\mathrm{reco}}}\xspace}
\newcommand{\mlb    }{\ensuremath{m_{lb}^{\mathrm{reco}}}\xspace}
\newcommand{\ptlb   }{\ensuremath{p_{\mathrm{T},lb}}\xspace}
\newcommand{\Rtt    }{\ensuremath{R_{3/2}}\xspace}
\newcommand{\dmt    }{\ensuremath{\Delta\mt}\xspace}
\newcommand{\mtpole }{\ensuremath{m_{\mathrm{top}}^{\mathrm{pole}}}\xspace}
\newcommand{\sig    }{\ensuremath{\sigma}\xspace}
\newcommand{\sqrts  }{\ensuremath{\sqrt{s}}\xspace}
\newcommand{\WWbar  }{\ensuremath{W^+W^-}\xspace}
\newcommand{\ttbar  }{\ensuremath{t\bar{t}}\xspace}
\newcommand{\ttbarjj}{\ensuremath{\ttbar\to\mbox{all-jets}}\xspace}
\newcommand{\ttbarll}{\ensuremath{\ttbar\to\mbox{dilepton}}\xspace}
\newcommand{\ttbarlj}{\ensuremath{\ttbar\to\mbox{lepton+jets}}\xspace}
\newcommand{\GeV    }{\ensuremath{\mathrm{GeV}}\xspace}
\newcommand{\TeV    }{\ensuremath{\mathrm{TeV}}\xspace}
\newcommand{\rhotot }{\ensuremath{\rho_{\mathrm{total}}}\xspace}
\newcommand{\Cpp    }{{\sc C++}\xspace}
\title{Measurements of the top quark mass with the ATLAS detector}
\ShortTitle{Measurements of the top quark mass with the ATLAS detector}
\author{
 \speaker{Richard Nisius}\thanks{On behalf of the ATLAS Collaboration}\\
 Max-Planck-Institut f\"ur Physik (Werner-Heisenberg-Institut)
 F\"ohringer Ring 6, D-80805 M\"unchen, Germany \\
 E-mail: \email{Richard.Nisius@mpp.mpg.de}
}
\abstract{The measurements of the top quark mass given are obtained from ATLAS
  data taken at proton--proton centre-of-mass energies of $\sqrt{s}=7$ and
  $8$~TeV.
 An extraction of the top quark pole mass~($m_{\mathrm{top}}^{\mathrm{pole}}$)
 at next-to-leading order~(NLO) is presented.
 This result is obtained from normalised differential cross-sections in the
 $t\bar{t}\to\mbox{dilepton}$ channel leading to:
 $m_{\mathrm{top}}^{\mathrm{pole}} = 173.2 \pm 0.9~(\mathrm{stat.}) \pm
 0.8~(\mathrm{syst.}) \pm 1.2~(\mathrm{theo.})$~GeV.
 In addition, measurements of $m_{\mathrm{top}}$ are discussed that are based on
 the template method performed in three $t\bar{t}$ decay channels.
 For all results the uncertainty is dominated by systematic effects.
 Finally, the 2016 ATLAS combined value of $m_{\mathrm{top}}$ is:
 $m_{\mathrm{top}}=172.84\pm0.34~(\mathrm{stat.})\pm0.61~(\mathrm{syst.})$~GeV,
 with a total uncertainty of 0.70~GeV, i.e.~a precision of 0.4$\%$.
}
\FullConference{
 The European Physical Society Conference on High Energy Physics\\ 
 5-12 July, 2017 \\ 
 Venice
}
\begin{document}
%
\section{Introduction}
\label{sec:intro}
 The Large Hadron Collider (LHC) is a top quark\footnote{Charge conjugation is
   implied throughout and natural units are used, $c=\hbar=1$.} factory.
 The largest rate of events with top quarks is obtained from
 \ttbar\ production. The \ttbar\ decay channels are classified by the
 \WWbar\ decays and are named the \ttbarll, \ttbarlj\ and \ttbarjj\ channels.

 The mass of the top quark~(\mt) is a fundamental parameter of the Standard
 Model~(SM) of particle physics.
 To obtain \mt\ from data, two conceptually different approaches are followed.
 Firstly, as has been done since the discovery of the top quark in 1995, \mt\ is
 measured from \ttbar\ final state objects by means of template analyses. The
 templates are obtained from Monte Carlo~(MC) simulated events using different
 assumed values for the top quark mass parameter in the program. Therefore,
 measurements of \mt\ obtained with this method relate to measurements of the
 input parameter of MC programs, i.e.~not to a specific program, since the
 differences of the programs used, are covered by the systematic uncertainty.
 Secondly, in recent years also \mtpole\ is extracted based on experimental
 quantities that are corrected for detector effects and compared to perturbative
 calculations at next-to-leading order~(NLO) in perturbative QCD, performed in a
 well-defined renormalisation scheme.
 Presently, the two attempts are complementary. While results of the first type
 are more precise, those of the second relate to a more accurate theoretical
 definition of the top quark mass.
 The relation between \mtpole\ and \mt\ is a matter of theoretical debate. Once
 this issue is resolved, analyses leading to the more precise results will be
 preferred.

 The details of the ATLAS~\cite{PERF-2007-01} analyses presented here are given
 in the respective
 publications~\cite{ATLAS-CONF-2017-044,TOPQ-2016-03,TOPQ-2013-02,TOPQ-2015-03}.
 In this short write-up, only the main aspects of the analyses are discussed.
%
\begin{figure*}[tbp!]
\centering
\subfigure[Data versus MC simulation]{%
  \includegraphics[width=0.42\textwidth]{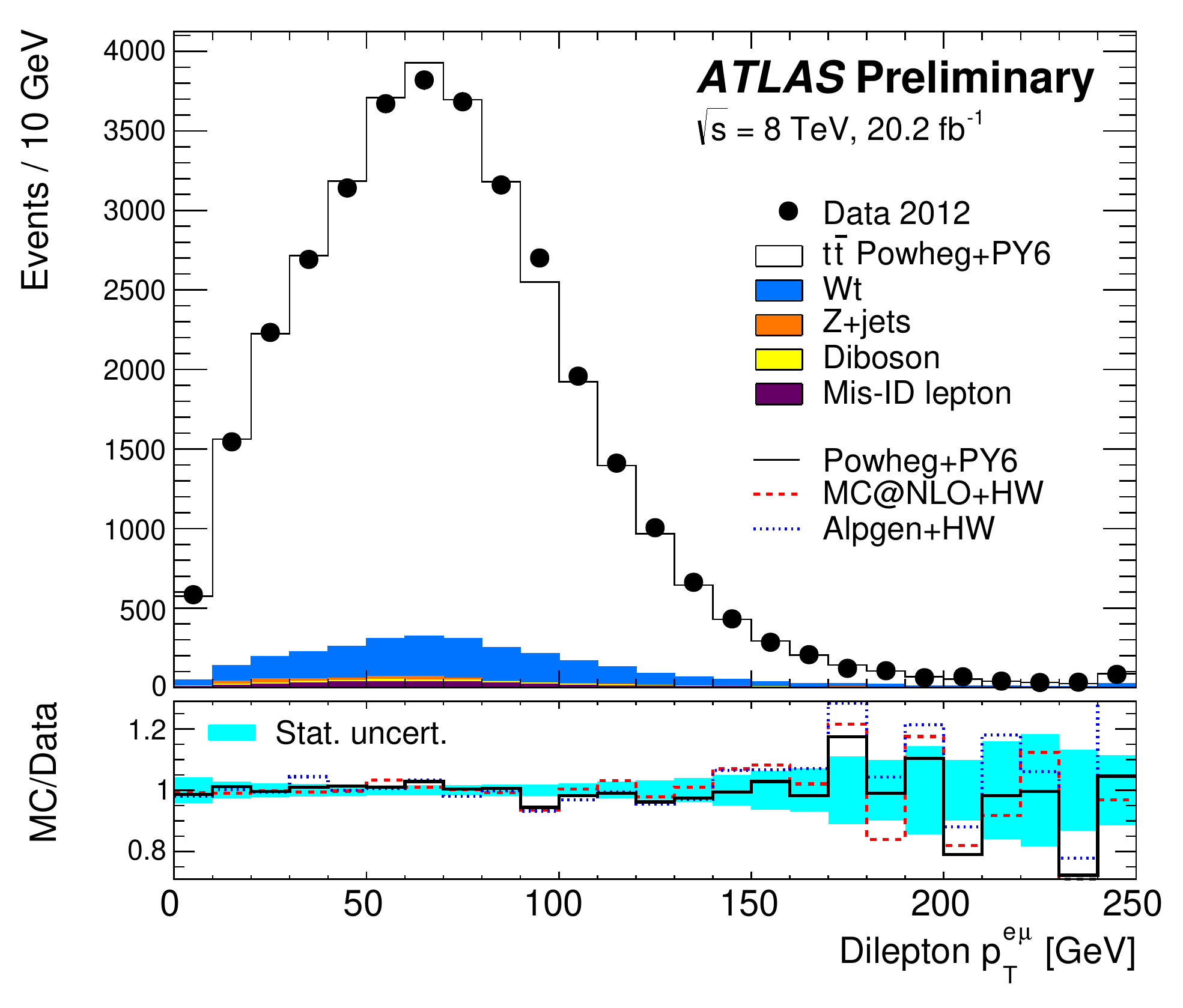}
  \label{fig:fig_01a}}
\subfigure[Data versus NLO MCFM prediction]{%
  \includegraphics[width=0.55\textwidth]{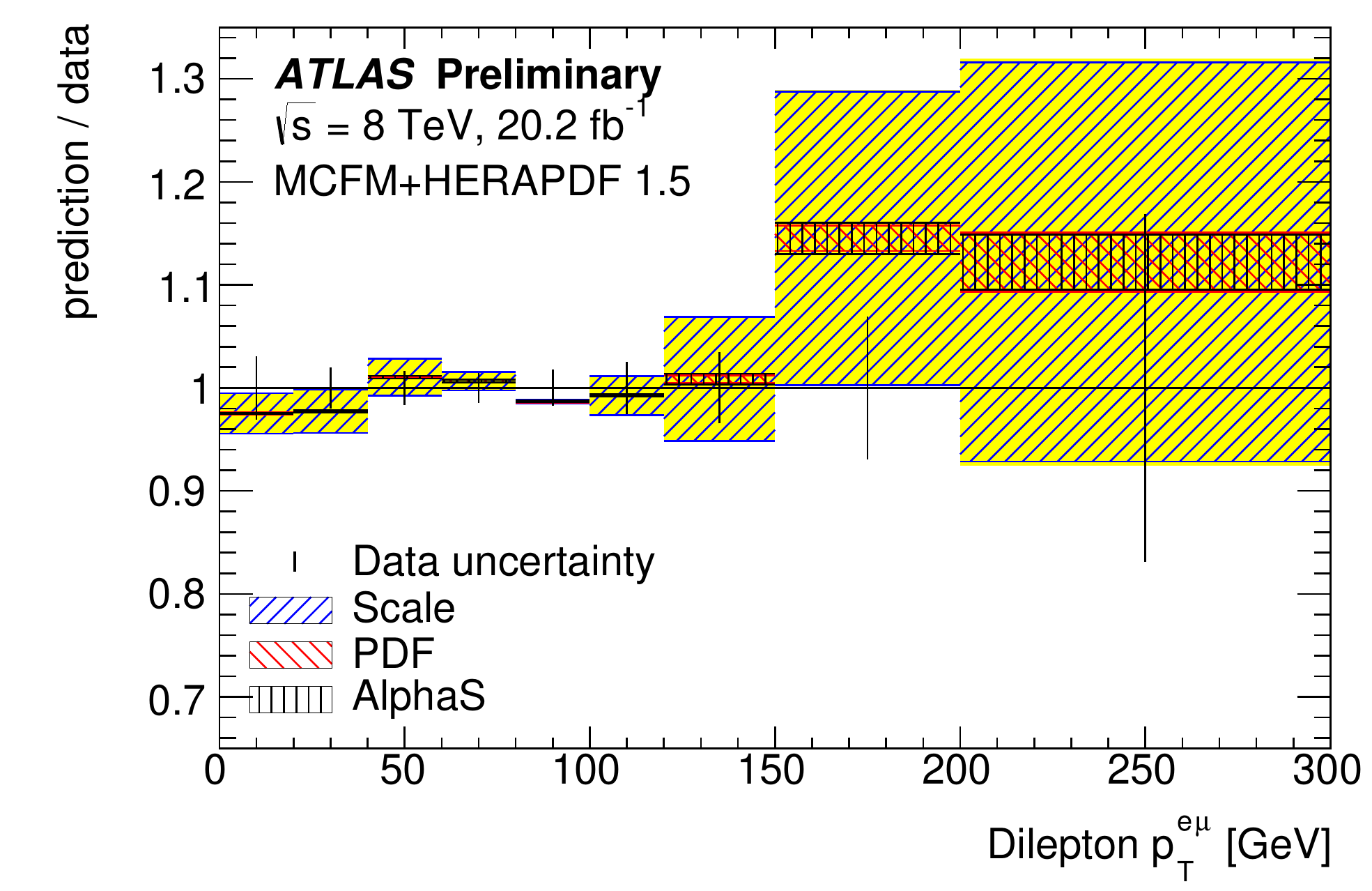}
  \label{fig:fig_01b}}
\hfill
\subfigure[Fitted values of \mtpole]{%
  \includegraphics[width=0.90\textwidth]{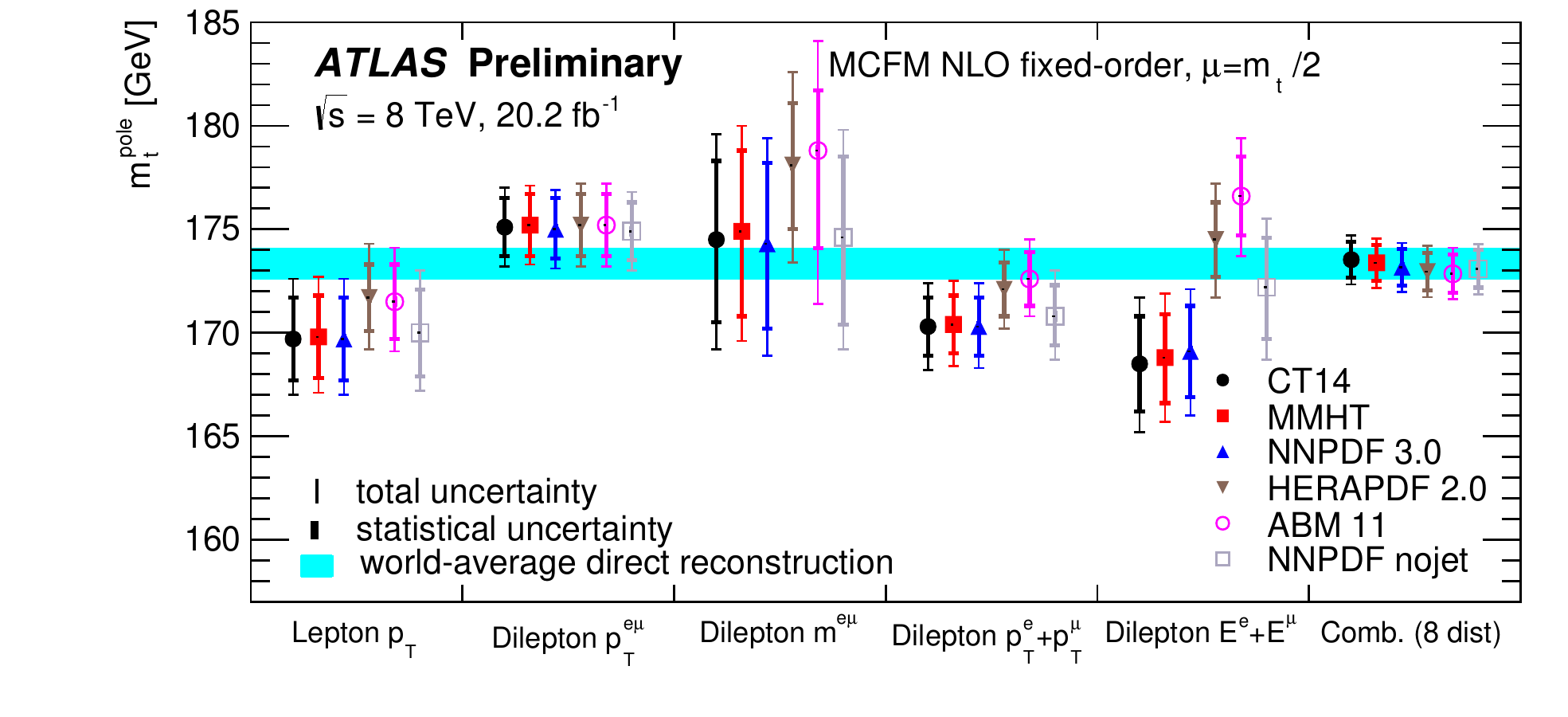}
  \label{fig:fig_01c}}
\caption{Extraction of \mtpole\ from normalised differential
  cross-sections~\protect\cite{ATLAS-CONF-2017-044}.
 Figure~(a) shows the distribution of the transverse momentum of the
 electron-muon system observed in data compared to MC simulation events.
 For this variable, figure~(b) shows the ratio of the predicted differential
 cross-section at NLO in production and decay and the measured one with
 experimental and theoretical uncertainties.
 Finally, figure~(c) shows \mtpole\ extracted from various variables and from a
 simultaneous fit to all of them. All results are given as a function of the
 proton PDF used in the extraction.
\label{fig:mtpole}}
\end{figure*}
%
\section{Determination of the top quark pole mass}
\label{sec:mpole}
 Because measurements of \mt\ are mostly limited by systematic uncertainties
 related to the hadronic final state, purely leptonic variables have been
 advertised for measurements of the top quark mass.
 The extraction~\cite{ATLAS-CONF-2017-044} of \mtpole\ is performed in the
 $\ttbar \to e\mu + X$ channel, using normalised lepton differential
 cross-sections, $\frac{1}{\sigma_x}\frac{d\sigma_x}{dx}$ for five variables,
 $x=p_{\mathrm T}^{\ell}, p_{\mathrm T}^{e\mu}, m^{e\mu}, p_{\mathrm
   T}^{e}$+$p_{\mathrm T}^{\mu}$ and $E^{e}$+$E^{\mu}$.
 One example distribution is shown in Fig.~\ref{fig:fig_01a}. Subsequently, the
 distributions are background subtracted, corrected to stable particle level
 using Powheg+Pythia6+CT10 as explained in Ref.~\cite{ATLAS-CONF-2017-044}, and
 finally normalised to unity.
 A clear sensitivity to the top quark mass is observed for all five variables.
 A comparison to a fixed order prediction at NLO in production and decay is
 shown in Fig.~\ref{fig:fig_01b}, where the bars denote the full data
 uncertainty.
 Within the sizeable uncertainties, the fixed order prediction describes the
 data.
 The measurements of \mtpole\ in Fig.~\ref{fig:fig_01c} are obtained from fits
 to individual differential cross-sections, or to all, including those for
 $\vert \eta^{\ell}\vert, \vert y^{e\mu}\vert, \Delta\Phi^{e\mu}$.
 The latter distributions are not sensitive to \mtpole, but help constrain the
 PDF and the renormalisation and factorisation scale variation induced
 uncertainties in \mtpole.
 The spread in the individual results of \mtpole\ is about 6~\GeV.
 The combined result is:
 $\mtpole = 173.2 \pm 0.9~(\mathrm{stat.})  \pm 0.8~(\mathrm{syst.})  \pm
 1.2~(\mathrm{theo.})~\GeV = 173.2 \pm 1.6$~\GeV.
 The theoretical uncertainty originates from PDF~(0.3~\GeV) and scale
 variations~(1.1.~\GeV), the latter obtained from variations using either fixed
 or dynamic~(e.g.~$E_{\mathrm{T}}/2$) scales.
 The first NLO extraction of \mtpole\ using this method results in an
 uncertainty of 1.6~\GeV, to be compared with the precision of the results
 presented in the next section.
%
\section{Measurement of the top quark mass}
\label{sec:monte}
 The results listed in Tab.~\ref{tab:tab_01} are obtained in the dilepton
 channel~\cite{TOPQ-2016-03}, the lepton+jets channel~\cite{TOPQ-2013-02} and
 finally, in the all-jets channel~\cite{TOPQ-2015-03}.
%
\begin{table*}[tb!]
\small
\begin{center} 
\begin{tabular}{rccccccc}
 Channel~(\sqrts) &
 Value & Statistics & Modelling & Background & Experimental & Total &
 Ref.\\\hline
 Dilepton~(8~\TeV) &
 172.99 &0.41 &$0.35\pm0.09$ &$0.08\pm0.01$ &$0.64\pm 0.04$ & 0.84 &
 \protect\cite{TOPQ-2016-03}\\
 Lepton+jets~(7~\TeV) &
 172.33 &0.75 &$0.53\pm0.11$ &$0.31\pm0.00$ &$0.82\pm 0.08$ & 1.27 &
 \protect\cite{TOPQ-2013-02}\\
 All-jets~(8~\TeV) &
 173.72 &0.55 &$0.70\pm0.16$ &         0.19 &$0.71\pm 0.04$ & 1.15 &
 \protect\cite{TOPQ-2015-03} \\
\hline
\end{tabular}
\end{center}
\caption{The measured values of \mt\ are given together with the statistical
  uncertainties, systematic uncertainties subdivided into modelling, background
  related and experimental uncertainties and the total uncertainty~(in [GeV]).
 For each systematic uncertainty listed, the first value corresponds to the
 uncertainty in \mt, and the second (if available) to the statistical precision
 of this uncertainty.
\label{tab:tab_01}}
\end{table*}
%
 In general, measurements of \mt\ receive large uncertainties induced by the
 jet energy scale~(JES) uncertainties, evaluated for all jets and by the
 relative $b$-to-light-jet energy scale~(bJES) uncertainty that only relates to
 $b$-quark initiated jets.
 Those uncertainties account for a large fraction of the experimental
 uncertainties listed in Tab.~\ref{tab:tab_01}.
 Aiming at the most precise value of \mt\ in the combination of results, and not
 in individual channels, methods are preferred that reduce the uncertainties,
 while retaining or even better, reducing the correlations~($\rho$) of the
 various estimators.
 This is because in this case large improvements over the most precise result,
 i.e.~the knowledge in \mt\ without combination, are obtained~\cite{NIS-1401}.
 In ATLAS, two main paths to mitigate the jet energy scale induced uncertainties
 in \mt\ are followed.
 The first is to use an \mt\ sensitive observable that is stabilised against JES
 variations.
 The second is to use additional information in the data sensitive to global
 changes in the jet energy scales.
 This information is then used for protecting the \mt\ sensitive observable
 against global shifts, by absorbing them into global jet energy scale factors,
 named JSF and bJSF.

\begin{figure*}[tbp!]
\centering
\subfigure[\mt\ sensitive distribution, all-jets channel]{%
  \includegraphics[width=0.42\textwidth]{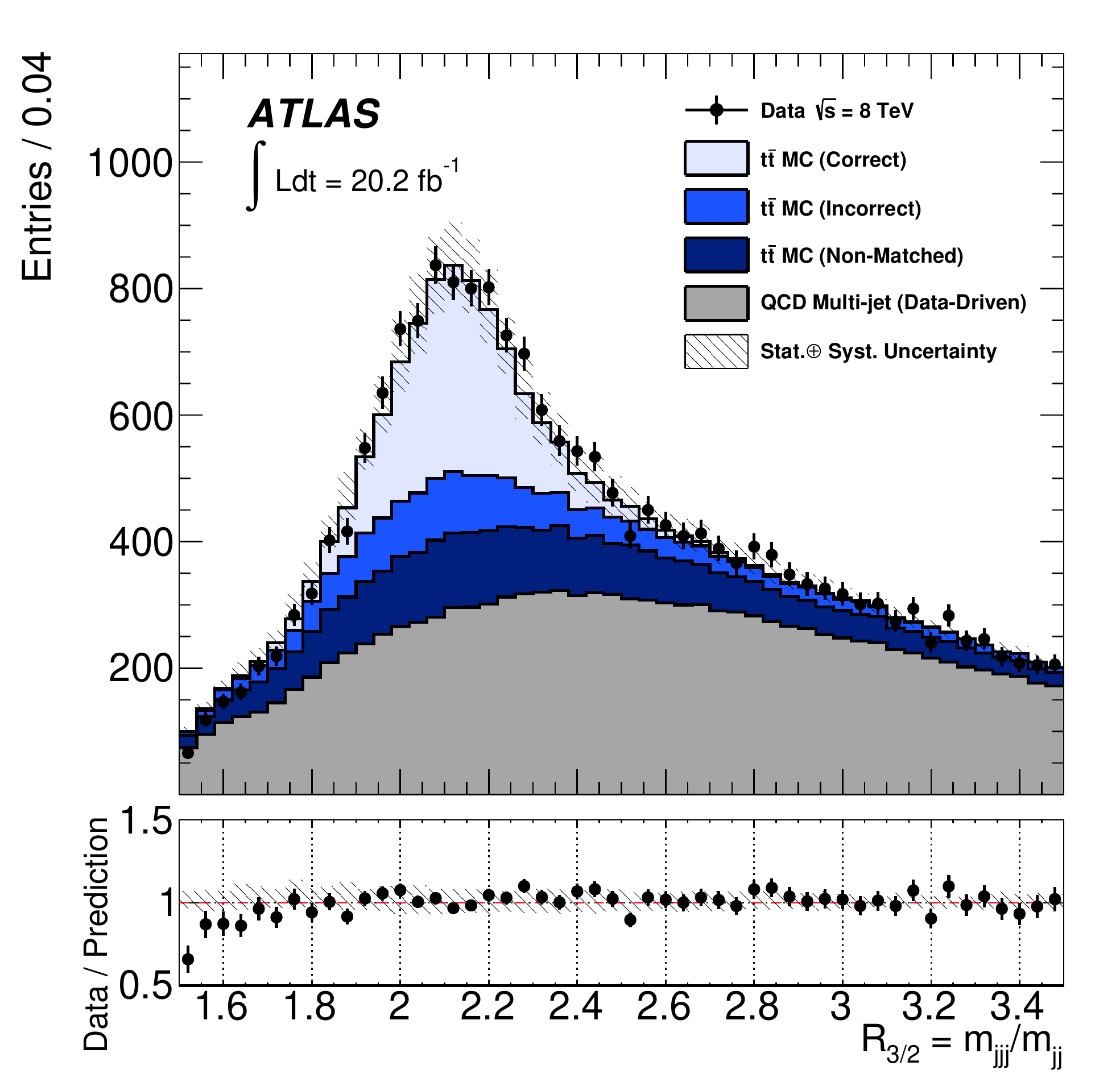}
  \label{fig:fig_02a}}
\subfigure[\mt\ sensitive distribution, lepton+jets channel]{%
  \includegraphics[width=0.55\textwidth]{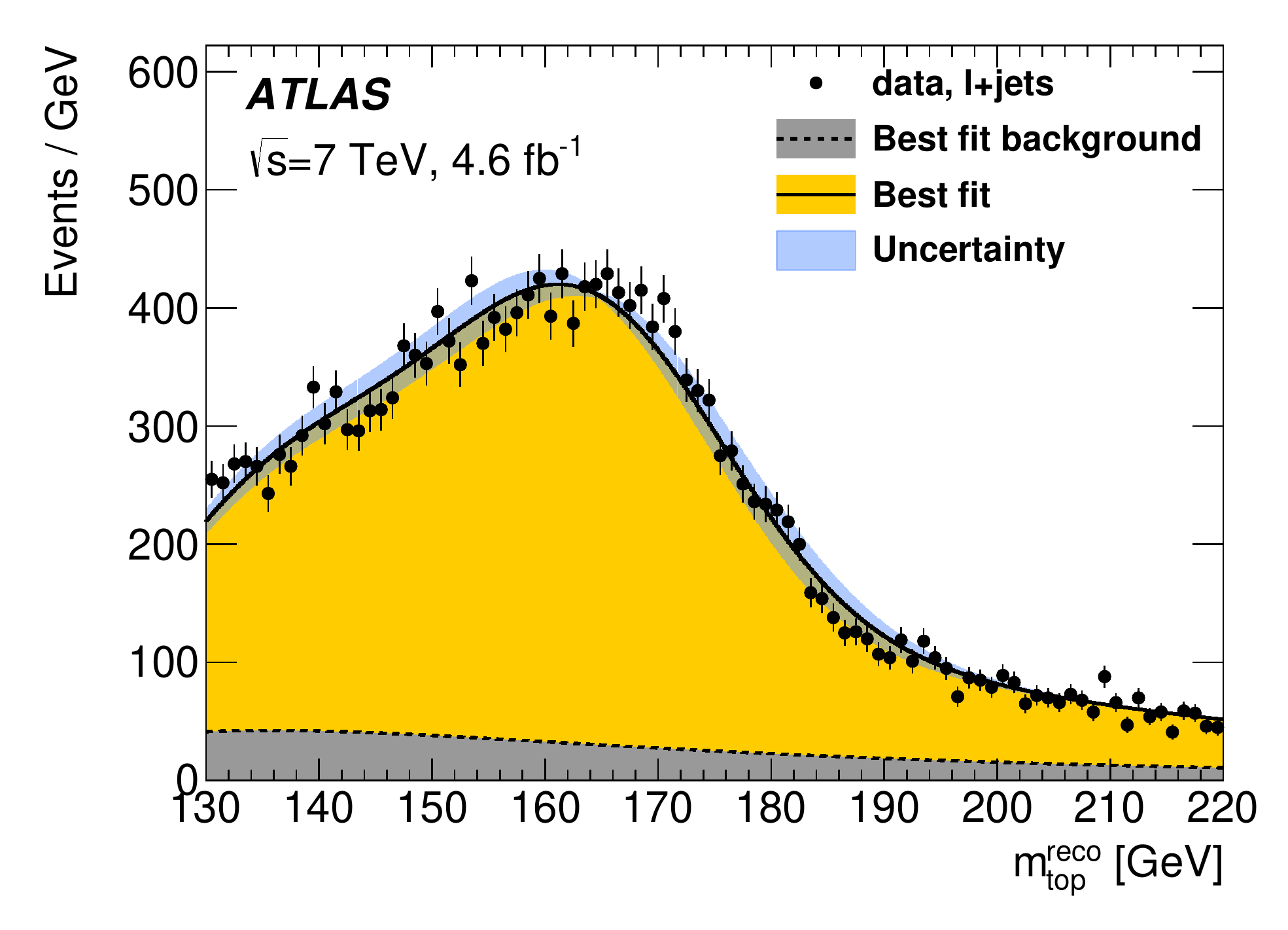}
  \label{fig:fig_02b}}
\hfill
\subfigure[\mt\ sensitive distribution, dilepton channel]{%
  \includegraphics[width=0.51\textwidth]{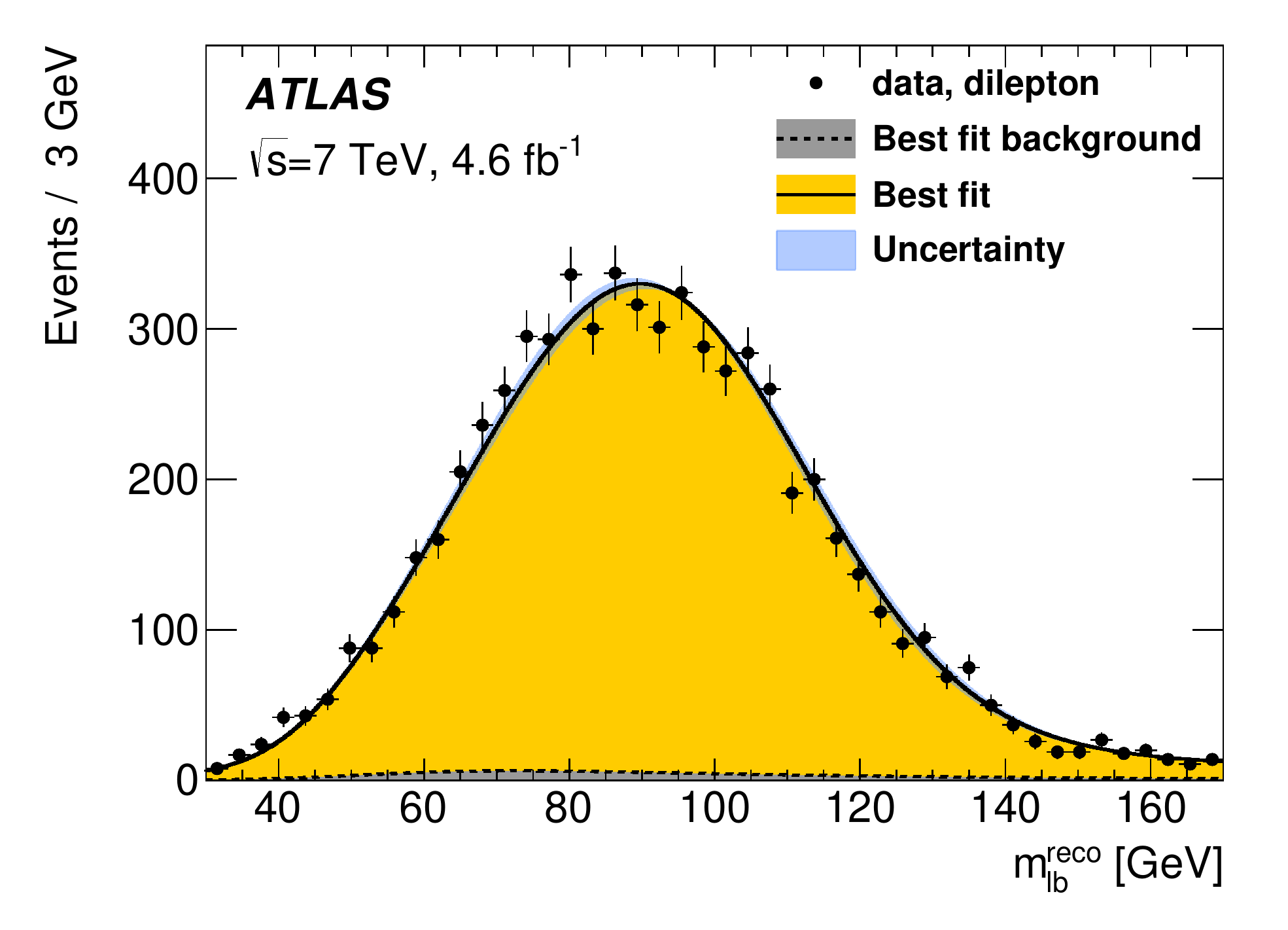}
  \label{fig:fig_02c}}
\subfigure[Uncertainty optimisation, dilepton channel]{%
  \includegraphics[width=0.47\textwidth]{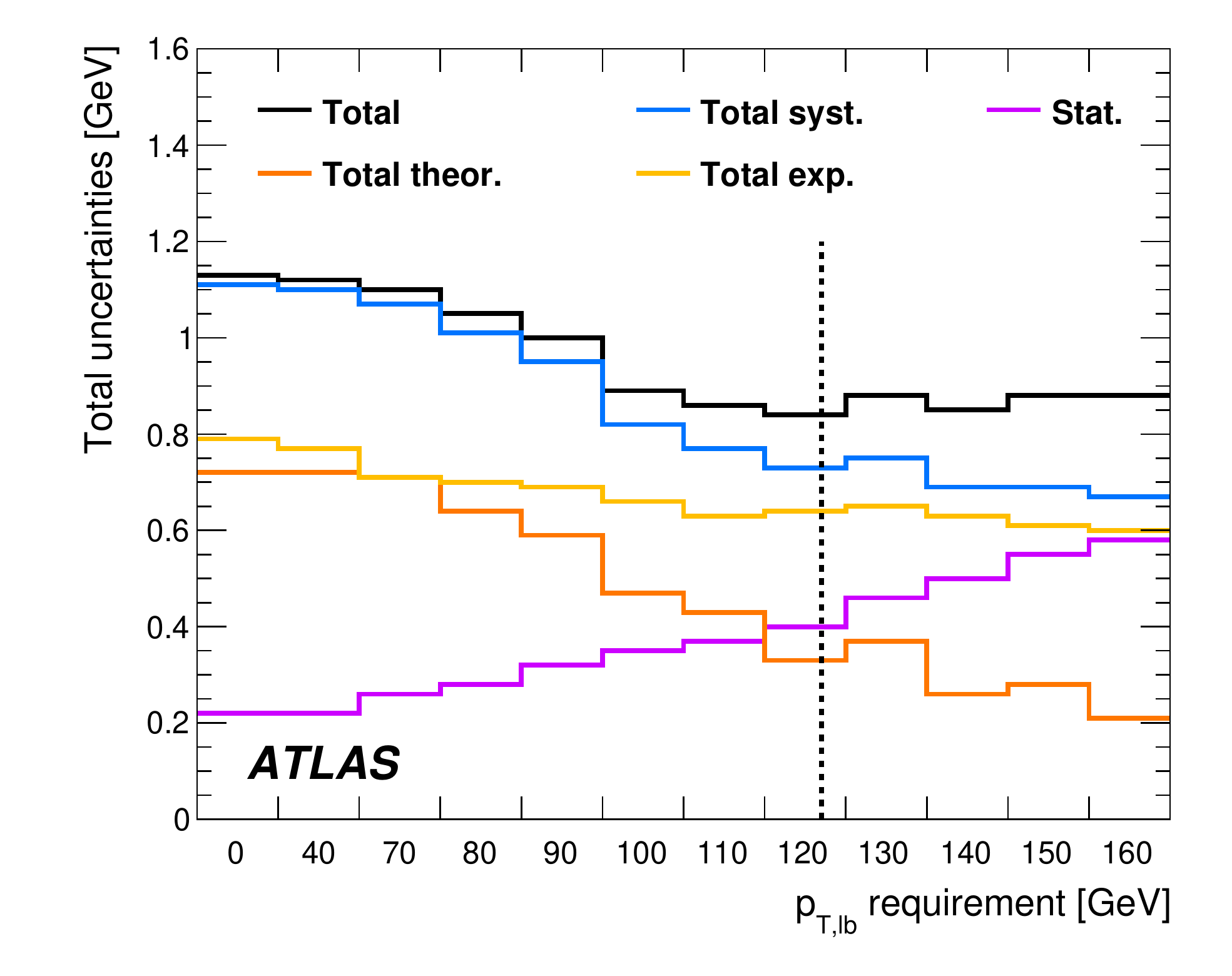}
  \label{fig:fig_02d}}
\caption{Measurements of \mt\ from template fits.
 Figure~(a) shows the ratio of the three-jet invariant mass to the two-jet
 invariant mass~(\Rtt) in the all-jets channel~\protect\cite{TOPQ-2015-03}.
 Figure~(b) shows the reconstructed top quark mass (\mtr) in the lepton+jets
 channel~\protect\cite{TOPQ-2013-02}.
 Figure~(c, d) are obtained from the two charged-lepton--$b$-jet systems in the
 dilepton channel~\protect\cite{TOPQ-2016-03}.
 Figure~(c) shows the average reconstructed invariant mass (\mlb) of the two
 systems.
 Finally, figure~(d) shows the total uncertainty~(\sig) in \mt\ for the dilepton
 channel as a function of the minimum requirement on the average transverse
 momenta of the two systems~(\ptlb).
\label{fig:fig_02}}
\end{figure*}

 The analysis in the all-jets channel uses the first path pioneered for the
 lepton+jets channel in Ref.~\cite{TOPQ-2011-15}. The analysis exploits the
 ratio of the three-jet mass and the two-jet mass~(\Rtt) as \mt\ sensitive
 variable, shown in Fig.~\ref{fig:fig_02a}.
 Since the JES induced variations in the jet energies at the same time apply to
 the jets in the numerator and denominator, most of the effects cancel.
 This cancellation results in a reduced uncertainty in \mt\ from this source
 than otherwise would be obtained when directly using the reconstructed top
 quark mass~(\mtr) for measuring \mt.
%

 The measurement in the lepton+jets channel uses the second path.
 Together with \mtr, shown in Fig.~\ref{fig:fig_02b}, two additional
 distributions are exploited.
 These are the reconstructed invariant mass of the $W$-boson and the ratio of
 transverse momenta of the two $b$-jets and the two light jets assigned to the
 hadronic $W$-boson decay.
 The first distribution is sensitive to the JES. The second distribution is
 sensitive to the bJES, while, as for $R_{3/2}$, the JES dependence mostly
 cancels in this ratio of transverse momenta.
 The price to pay is additional contributions to the statistical uncertainty in
 \mt\ caused by fitting to more distributions.
 However, for sufficiently large data samples, this loss in statistical
 precision is more than compensated for by the strong reduction in the
 respective systematic uncertainty in \mt, induced by the jet energy scale
 uncertainty.
 This situation is realised in Ref.~\cite{TOPQ-2013-02}.
 Not only does this achieve a smaller total uncertainty~(\sig) in \mt\ for this
 decay channel, it also transforms a significant part of the systematic
 uncertainty into a statistical uncertainty.
 This has two positive consequences. Firstly, this statistical component to the
 uncertainty will naturally be reduced by including more data. 
 Secondly, given that the results in \mt\ for the various decay channels are
 uncorrelated with respect to their statistical uncertainties, it also
 potentially helps in reducing the estimator correlations.
%
\begin{figure*}[tbp!]
\centering
\subfigure[Pairwise correlation - lepton+jets versus dilepton]{%
  \includegraphics[width=0.49\textwidth]{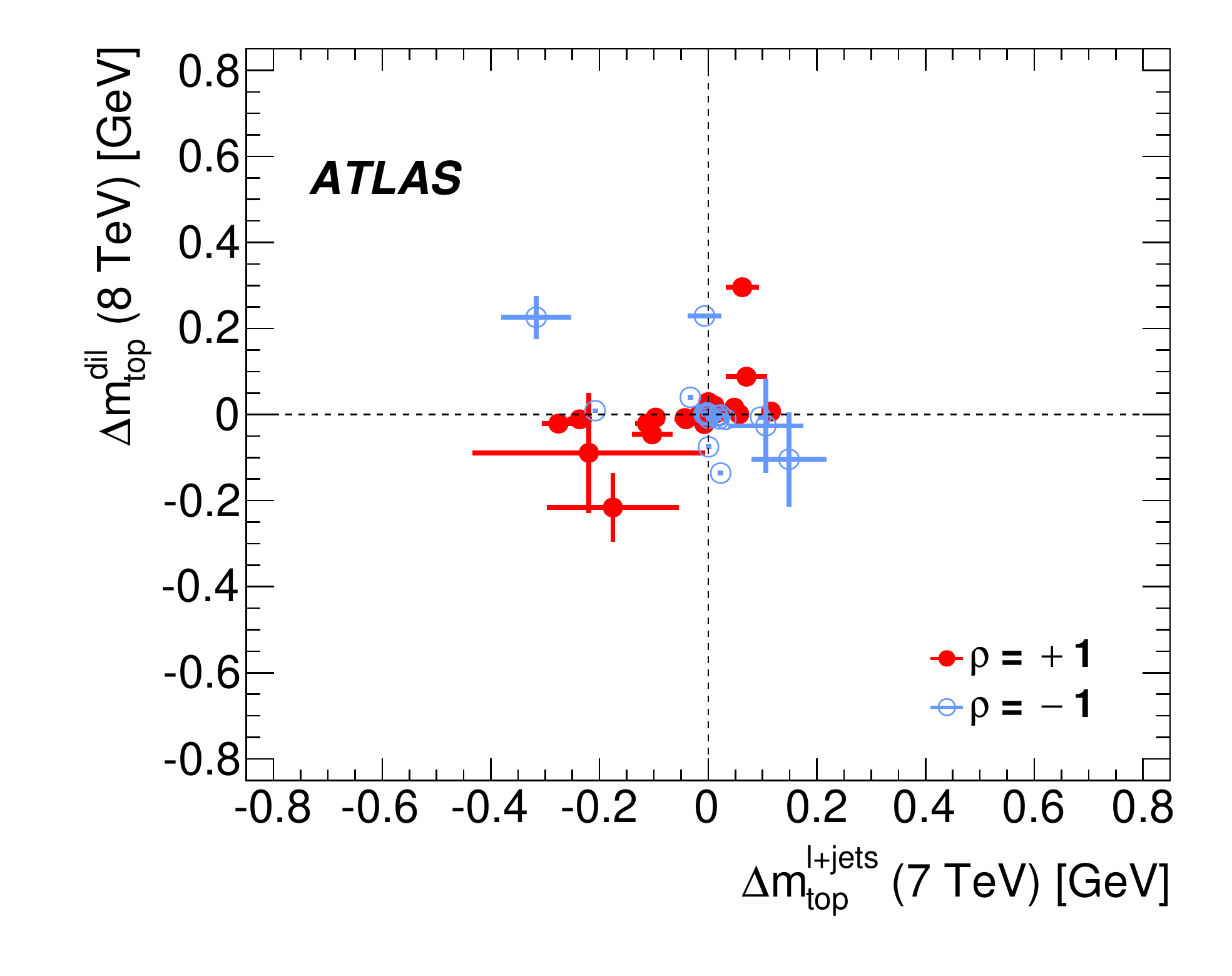}
  \label{fig:fig_03a}}
\subfigure[Combination of the two most precise results]{%
  \includegraphics[width=0.46\textwidth]{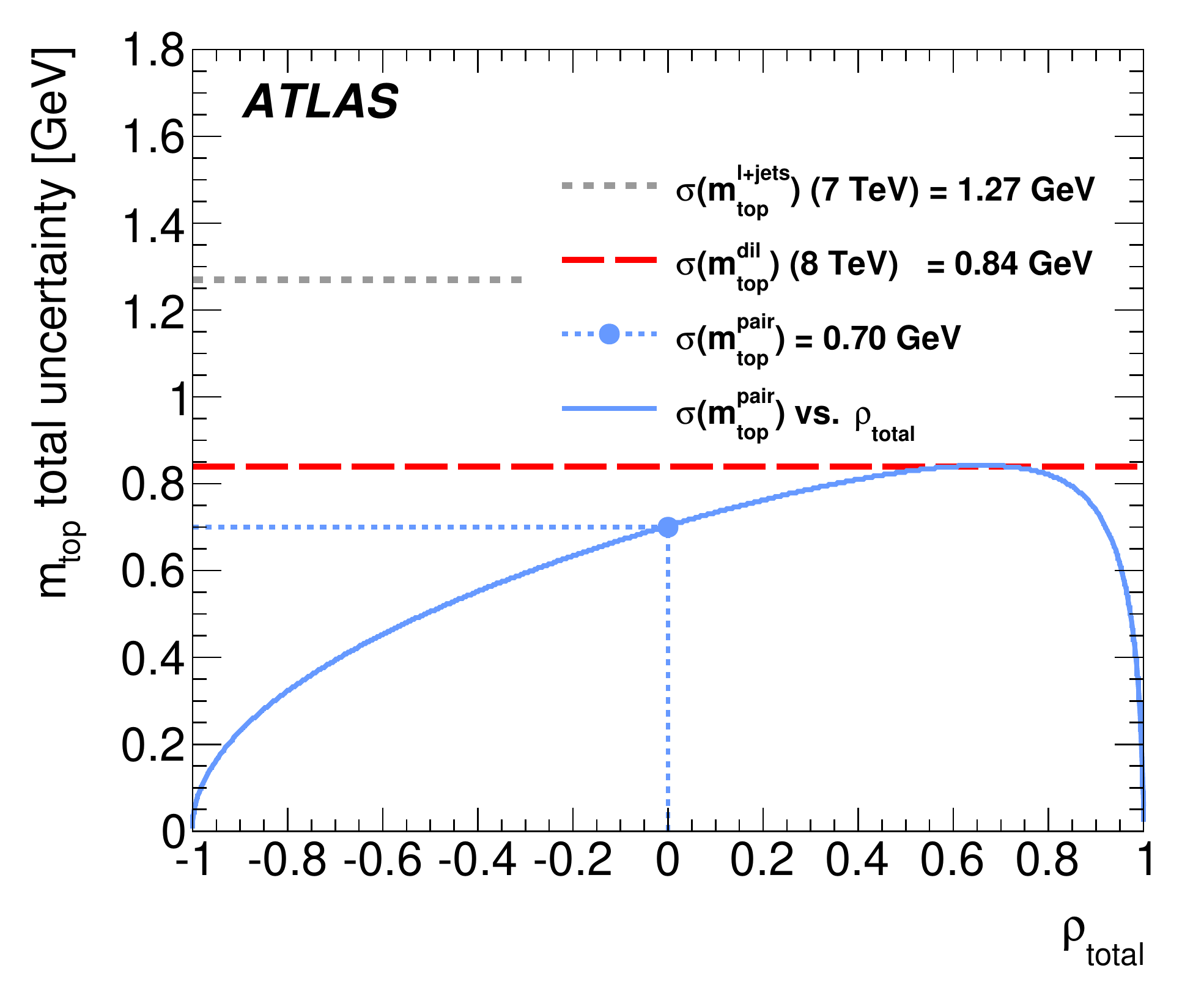}
  \label{fig:fig_03b}}
\caption{Figure~(a) shows the differences in the top quark mass~(\dmt) when
  simultaneously varying a pair of measurements for the subcomponents of a
  systematic uncertainty using the pairs indicated~\protect\cite{TOPQ-2016-03}.
 The red full points indicate the fully correlated cases, the blue open points
 the anti-correlated ones. See text for further details.
 Figure~(b) shows the uncertainty in the combined result of the combination of
 the most precise pair of results as a function of the estimator correlation
 \rhotot.
 The blue point corresponds to the actual estimator
 correlation~\protect\cite{TOPQ-2016-03}.
\label{fig:fig_03}}
\end{figure*}

 The missing hadronic $W$-boson decay in the dilepton channel prevents the use
 of the above methods.
 However, the clean final state in this channel results in high selection
 efficiency at high signal purity. 
 The low background is barely visible for the distribution of the average
 invariant mass of the charged-lepton--$b$-jet systems~(\mlb) shown in
 Fig.~\ref{fig:fig_02c}.
 This would result in a very unbalanced composition of statistical and
 systematic uncertainty, i.e.~indicating a non-optimised analysis.
 Consequently, the large data sample is used to reduce the total uncertainty by
 trading statistical for systematic precision based on additional phase space
 restrictions.
 The result of this optimisation~\cite{TOPQ-2016-03} as a function of the
 average transverse momentum of the two charged-lepton--$b$-jet systems~(\ptlb)
 is shown in Fig.~\ref{fig:fig_02d}.
 Compared to the result without this requirement, a reduction of $26\%$ in the
 uncertainty in \mt\ is achieved. This requirement removes $74\%$ of the
 original events, thereby resulting in an $86\%$ increase in the statistical
 uncertainty, i.e.~no gain in resolution is achieved by this phase space
 restriction.
 Again, given a significant part of the systematic uncertainty is transformed
 into a statistical uncertainty, this potentially helps in reducing the
 estimator correlations.
%
\section{Combination of measurements}
\label{sec:combi}
 The combination is performed using the best linear unbiased estimate~(BLUE)
 method~\cite{VAL-0301,NIS-1401} in a \Cpp\ implementation described in
 Ref.~\cite{NIS-1301}.
 The BLUE method combines measurements based on a linear combination of the
 inputs.
 The coefficients (BLUE weights) are determined via the minimisation of the
 total variance of the combined result. They can be used to construct measures
 for the importance of a given single measurement in the
 combination~\cite{NIS-1401}.
 The measured values of \mt, the list of uncertainty components and the
 correlations of the estimators for each uncertainty component have to be
 provided.
 The first two are given in the respective publications, while the estimator
 correlations need to be obtained.
 As developed in Ref.~\cite{TOPQ-2013-02}, for the ATLAS combination of
 measurements of \mt, the correlations are evaluated for each source of
 systematic uncertainty as displayed in Fig.~\ref{fig:fig_03a} for the most
 precise pair of estimators.
 Each point corresponds to one subcomponent of the systematic uncertainty in
 \mt, obtained by varying this pair of measurements for this
 subcomponents. Using MC simulated events, pseudo-experiments are constructed of
 the size of the ATLAS data sample and \mt\ is fitted to those.
 The point is located at the mean values of the observed shifts in the top quark
 mass~(\dmt) for both analyses. This location is calculated from the
 pseudo-experiments.
 The cross indicates the statistical precisions in the systematic uncertainties
 as given by the precision of the pseudo-experiments.
 Uncertainties for which the estimators are correlated are located in the first
 and third quadrant and are shown as red full points, the anti-correlated cases,
 located in the remaining two quadrants, are displayed by open blue points.
 The ATLAS combination is dominated by the two input results shown in
 Fig.~\ref{fig:fig_03b}. At the quoted precision, the combination of just this
 pair achieves the identical total uncertainty as the full combination.
 Finally, the 2016 combined ATLAS result for the top quark mass is:
 $\mt=172.84\pm0.34~(\mathrm{stat.})\pm0.61~(\mathrm{syst.})~\GeV=172.84\pm0.70\,\,\GeV$,
 with a precision of 0.4$\%$.
 Evaluating statistical uncertainties for each systematic uncertainty in
 \mt\ allows for performing stability tests on the combination without ad-hoc
 assignments of variations in estimator correlations.
 For each combination, the size of the uncertainty and the correlation are
 newly evaluated, based on random variations of each systematic uncertainty
 within its statistical precision.
 As a result, both the combined value and the corresponding uncertainty are
 stable to within 0.03~\GeV.
%
\section{Conclusions}
\label{sec:concl}
 An extraction of the top quark pole mass at NLO is presented. The result is
 obtained from normalised differential cross-sections in the \ttbarll\ channel.
 The uncertainty in the result of: 
 $\mtpole=173.2\pm0.9~(\mathrm{stat.})\pm0.8~(\mathrm{syst.})\pm1.2~(\mathrm{theo.})~\GeV = 173.2 \pm 1.6\,\,\GeV$,
 is dominated by systematic effects, mainly by scale variations.

 In addition, measurements of \mt\ are discussed, which are based on the
 template method in the three \ttbar\ decay channels, \ttbarll, \ttbarlj\ and
 \ttbarjj.
 The uncertainties in the two most precise results in \mt\ have been reduced
 using methods paying attention to the correlations of the estimators, enabling
 a significant gain in precision in \mt\ in their combination.
 The 2016 combined ATLAS result for the top quark mass is:
 $$\mt=172.84\pm0.34~(\mathrm{stat.})\pm0.61~(\mathrm{syst.})~\GeV=172.84\pm0.70\,\,\GeV,$$
 with a precision of 0.4$\%$.
 The statistical precision in the total uncertainty is 0.03~\GeV.
%
\providecommand{\href}[2]{#2}\begingroup\raggedright\endgroup
%
\end{document}